\documentstyle[prl,aps,epsfig,twocolumn]{revtex}

\begin{document}

\title{Two-Kaon Correlations in Central Pb + Pb Collisions at 158~A~GeV/{\it c}}

\author{
}

\twocolumn[
\begin{center}
{\large\bf Two-Kaon Correlations in Central Pb + Pb Collisions at 158~A~GeV/{\it c}}\\
I.G.~Bearden,$^{1}$ 
H.~B\o ggild,$^{1}$ 
J.~Boissevain,$^{2}$ 
P.H.L.~Christiansen,$^{1}$ 
L.~Conin,$^{3}$ 
J.~Dodd,$^{4}$
B.~Erazmus,$^{3}$ 
S.~Esumi,$^{5}$
C.W.~Fabjan,$^{6}$ 
D.~Ferenc,$^{7}$ 
A.~Franz,$^{6}$ 
J.J.~Gaardh\o je,$^{1}$ 
A.G.~Hansen,$^{1}$ 
O.~Hansen,$^{1}$ 
D.~Hardtke,$^{8}$
H. van~Hecke,$^{2}$ 
E.B.~Holzer,$^{6}$ 
T.J.~Humanic,$^{8}$ 
P.~Hummel,$^{6}$ 
B.V.~Jacak,$^{9}$ 
K.~Kaimi,$^{5\ast}$ 
M.~Kaneta,$^{5}$
T.~Kohama,$^{5}$ 
M.~Kopytine,$^{9\dagger}$ 
M.~Leltchouk,$^{4}$ 
A. Ljubi\v{c}i\'c Jr.,$^{7}$   
B.~L\"orstad,$^{10}$ 
N.~Maeda,$^{5}$ 
L.~Martin,$^{3}$ 
A.~Medvedev,$^{4}$ 
M.~Murray,$^{11}$ 
H.~Ohnishi,$^{5}$          
G.~Pai\'c,$^{6,8}$ 
S.U.~Pandey,$^{8}$
F.~Piuz,$^{6}$ 
J.~Pluta,$^{3}$ 
V.~Polychronakos,$^{12}$ 
M.~Potekhin,$^{4}$ 
G.~Poulard,$^{6}$ 
D.~Reichhold,$^{8}$
A.~Sakaguchi,$^{5}$ 
J.~Schmidt-S\o rensen,$^{10}$ 
J.~Simon-Gillo,$^{2}$
W.~Sondheim,$^{2}$ 
T.~Sugitate,$^{5}$ 
J.P.~Sullivan,$^{2}$ 
Y.~Sumi,$^{5}$
W.J.~Willis,$^{4}$ 
K.~Wolf,$^{11\ast}$  
N.~Xu,$^{2}$ and 
D.S. Zachary$^{8}$ \\
(NA44 Collaboration)\\

{\it
$^{1}$Niels Bohr Institute, DK-2100 Copenhagen, Denmark \\
$^{2}$Los Alamos National Laboratory, Los Alamos, NM 87545 \\
$^{3}$Nuclear Physics Laboratory of Nantes, 44072 Nantes, France \\
$^{4}$Department of Physics, Columbia University, New York, NY 10027 \\
$^{5}$Hiroshima University, Higashi-Hiroshima 739-8526, Japan \\
$^{6}$CERN, CH-1211 Geneva 23, Switzerland \\
$^{7}$Rudjer Bo\v{s}kovi\'c Institute, Zagreb, Croatia \\
$^{8}$Department of Physics, The Ohio State University, Columbus, OH 43210 \\
$^{9}$State University of New York, Stony Brook, NY 11794 \\
$^{10}$Department of Physics, University of Lund, S-22362 Lund, Sweden \\
$^{11}$Cyclotron Institute, Texas A\&M University, College Station, TX 77843 \\
$^{12}$Brookhaven National Laboratory, Upton, NY 11973 \\
}
(published in volume 87, issue 11 of Physical Review Letters with
electronic identifier 112301) 
\end{center}
\vspace{-0.2in}
\newlength{\leftindent}
\newlength{\rightindent}
\newlength{\minipagewidth}
\leftindent=1.0in
\rightindent=1.0in
\minipagewidth=\textwidth
\addtolength{\minipagewidth}{-\leftindent}
\addtolength{\minipagewidth}{-\rightindent}

\noindent
\hspace*{\leftindent}
\center
\begin{minipage}{\minipagewidth}
Two-particle interferometry of positive kaons is studied in Pb + Pb
collisions at mean transverse momenta $<p_{T}>\approx 0.25$ and  
0.91 GeV/{\it c}. A three-dimensional analysis was applied to the 
lower $p_T$ data, while a two-dimensional analysis was used for the 
higher $p_T$ data. We find that the source-size parameters are 
consistent with the $m_T$ scaling curve observed in pion-correlation 
measurements in the same collisions, and that the duration time of 
kaon emission is consistent with zero within the experimental
sensitivity.

\rightline{PACS Numbers: 25.75.Gz}

\quad
\end{minipage}
]

\narrowtext

Experimental studies of high-energy nuclear collisions at the BNL-AGS 
and CERN-SPS accelerators (beam energies from 10 to 200 GeV/nucleon) 
have revealed interesting features of hot and dense nuclear matter, 
and some characteristic signatures of a quark-gluon-plasma (QGP) phase 
have been reported~\cite{QM99}. If the hadronic source is formed in a 
first-order phase transition from a QGP phase in the course of collision, 
the hadronic expansion may slow down due to a softening of the equation 
of state. In such a case, a long duration time of particle emission is 
anticipated. 
Since a finite duration of particle emission increases the effective 
source size in the direction of particle velocity, and since the 
shape of the two-particle correlation function is related to the 
effective source size, a difference between the widths of the peaks 
in the correlation functions in the direction of pair (``outward'') 
and perpendicular to it (``sideward'') might be a signature of 
QGP~\cite{Bertsch89}.
At SPS energies, systematic studies 
of particle correlations have been performed from $p + A$ to Pb + Pb 
collisions~\cite{NA35,NA44,PRC(John),WA98,NA49}. 
From the pion correlation studies in the Pb + Pb collisions, 
NA44~\cite{PRC(John)} reported that the two transverse radius 
parameters in the outward and sideward directions in the longitudinal
center of mass system (LCMS) frame 
(see below) are similar, implying no long duration time of emission. 
WA98~\cite{WA98} measured the correlation function with the 
generalized Yano-Koonin parametrization and found the $R_{0}$ 
parameter, which reflects the duration of emission, is compatible with zero. 
While NA49~\cite{NA49} observed a finite $R_{0}$ parameter in the 
Yano-Koonin-Podgoretskii parametrization to be approximately 
3--4 fm/{\it c}, the value is small and not consistent with what 
would be expected from a strong first-order phase transition. 
All the experimental data support no long-lived intermediate 
hadron-parton mixed phase during the pion emission.  
There are discussions, however, that the pion correlation functions 
might be distorted due to a large amount of decays from long-lived 
resonances, while kaon measurements can serve as a more sensitive 
probe of the space-time evolution~\cite{Gyulassy90,Bolz93,Wiedemann97}. 
The kaon duration time was measured in S + Pb collisions, 
but a long-lived mixed phase was also excluded by the data~\cite{NA44}.  
In this letter, we present the first results of $K^+K^+$ correlations 
in central Pb + Pb collisions at the SPS energy and extract 
the duration time of the kaon emission.

The data were taken at the CERN-SPS with a 158 GeV/{\it c} per nucleon 
lead-ion beam incident on a lead target, using the NA44 spectrometer 
at two laboratory angles: 44 and 131 mrad with respect to the beam 
axis.
A description of the NA44 apparatus can be found 
elsewhere~\cite{NA44,PRC(John)}.
The lead beam was transported in vacuum up to a beam counter. 
The signal amplitude from the beam counter ensures a single lead 
ion impinging on the lead target at the entrance edge of the first 
dipole magnet. Behind the target, there were two multiplicity detectors 
characterizing each event. 
A set of scintillator bars was used primarily to generate a centrality 
trigger by measuring charged secondaries. A silicon pad detector with 
a full coverage in the azimuthal direction and segmented into 512 pads 
measured the $dN/d\eta$ distribution of charged particles in the 
pseudorapidity range between 1.5 and 3.3.

The secondary particles were transported to the tracking section 
through another dipole magnet and three quadrupole magnets. 
The quadrupole magnets control the three-dimensional momentum 
acceptance of the spectrometer. Two sets of quadrupole settings, 
referred to as the ``horizontal'' and ``vertical'' focus settings, 
were used. The horizontal (vertical) setting favors a wide momentum 
acceptance in the horizontal (vertical) space $p_x$ ($p_y$), and 
reduces the acceptance $p_y$ ($p_x$). 
In the discussion which follows, the direction with wide acceptance
for a given mode (e.g., $p_x$ for the horizontal mode) is referred 
to as the ``favored'' direction and the direction with narrow acceptance 
(e.g., $p_y$ for the horizontal mode) is called ``unfavored''.
For the small angle measurements, collimators between the first dipole 
and the first quadrupole magnets further reduced the acceptance in the 
unfavored direction. These collimators reduced the number of particles 
in the acceptance without reducing the acceptance in the favored 
direction. The momentum settings of the spectrometer were 6 GeV/{\it c} 
for the small angle (low $p_T$) measurements and 7.5 GeV/{\it c} 
for the large angle (high $p_T$) measurements. 

\begin{figure}[t]
\label{pid}
\epsfig{file=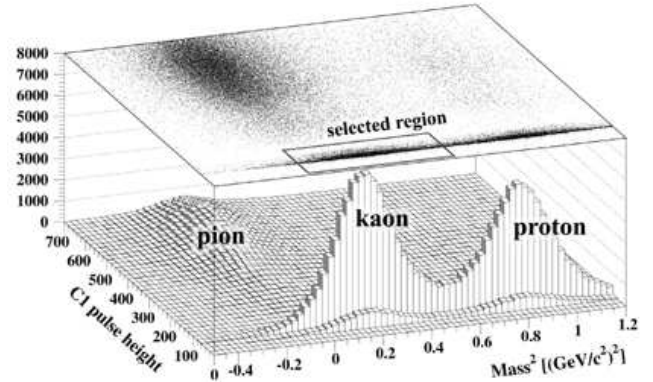,width=3.375in}
\caption{The particle identification using the mass-squared and the 
C1 pulse height. Kaon and proton peaks are seen near the C1 pedestal 
(70 ch.), while a bump at around 600 ch. on C1 and $Mass^2\approx 0$ 
is of pion events accumulated with a hardware scaled-down trigger. 
On the top plane, a scatter plot is shown with the PID selection 
adopted in this analysis. }
\label{fig:1}
\end{figure}

After the magnets, there were three tracking chambers (PC, SC1, SC2), 
three threshold-type gas Cherenkov counters (C1, C2, TIC) and three 
scintillator hodoscopes (H2, H3, H4). The pad chamber (PC) and 
the strip chambers (SC) provided precise hit position for the 
tracking algorithm. Although their 
position information was less precise, 
the hodoscopes were also used in track finding. 
C1 was the primary device for pion/kaon separation in the small angle 
measurements, while C2 and TIC were used for pion/kaon separation in 
the large angle measurements. 
Tracks were reconstructed by requiring hits 
in the tracking chambers and 
hodoscopes on straight lines. The three-dimensional momenta were 
calculated from the track information through a matrix, of which the 
elements were determined by a Monte Carlo simulation combined with 
the TURTLE code for the particle transportation through the magnetic field. 
Particle identification used time-of-flight information between the 
beam counter and the hodoscopes along with pulse height information 
from the Cherenkov counters. 
Figure~\ref{fig:1} shows an example of such a plot in the small angle 
measurements and also the particle-identification (PID) selection
adopted in this analysis.
Contaminations of pion and proton tracks in the kaon-identified 
pair-samples were evaluated to be 0.15\% and 3.9\%, respectively. 
Therefore, contaminations of $\pi\pi$ and $pp$ pairs in the $KK$ 
samples were negligibly small. 

At the small angle, the spectrometer covers the rapidity range of 
2.9--3.3 with a $p_T$ window below 0.6 GeV/{\it c} 
($<p_T>\approx 0.25$ GeV/{\it c}). 
The rapidity coverage in the large angle measurements is 2.4--2.9 with 
the $p_T$ window of 0.7--1.4 GeV/{\it c} 
($<p_T>\approx 0.91$ GeV/{\it c}). 
Note that the colliding system is symmetric and the central rapidity 
is at 2.9, so the rapidity coverage in the high and low $p_T$ 
measurements overlap well. 
The trigger required a lead ion in the beam counter, a large
number of secondary particles in the scintillator bars, and at
least two hits in hodoscopes H2 and H3. Off-line we required
at least two kaons in each event and found the centrality of
the lower $p_T$ data set to be around 10\% of the most central 
events while that of the high $p_T$ sample was 18\%.
      
The two-particle correlation function is defined by 
\begin{equation} 
C_{raw}(\vec{p}_1,\vec{p}_2)= {P_2(\vec{p}_1,\vec{p}_2)
	\over P_1(\vec{p}_1)P_1(\vec{p}_2)} 
        \approx {Real(\vec{p}_1,\vec{p}_2)
        \over Back(\vec{p}_1,\vec{p}_2)}, 
\label{eq:C_raw}
\end{equation} 
where the numerator is the joint probability of detecting two particles  
with momenta $\vec{p}_1$ and $\vec{p}_2$, 
while the denominator is the product of the probabilities of detecting 
single particles with momenta $\vec{p}_1$ and $\vec{p}_2$. 
The denominator was obtained by mixing two tracks picked up from 
two randomly selected different events.
Ten mixed background pairs 
were generated for each real pair to reduce statistical uncertainties. 
There still remain several effects, which affect the true correlation 
function. The limited acceptance of the spectrometer distorts the real 
two-particle spectrum, and the finite momentum resolution smears the peak 
in the correlation function. The repulsive Coulomb force separates 
particles of a pair going out at $\vec{p}_1 \simeq \vec{p}_2$ and 
strongly suppresses the yield at $\vec{q} = \vec{p}_1 - \vec{p}_2 \sim 0$. 
The background two-particle spectrum generated by the event mixing 
method still contains effects of two-particle correlations, since 
a particle in a real event is always accompanied by another particle 
nearby in phase space. Since the degree of these effects depends on 
the strength of particle correlations of interest in a source, 
these effects were corrected through a Monte Carlo (MC) based 
iteration method. A $C_{2}$ function (described in the next 
paragraph) was assumed for a given set of source-size parameters ($R$'s and 
$\lambda$), and MC events were generated taking the spectrometer 
response into account. The wave integration method~\cite{Pratt86} 
was employed to simulate Coulomb effects in the finite source volume. 
The MC events were analyzed using exactly the same procedure 
as was applied to the experimental data. The correction factors to 
the $C_{raw}$ function were evaluated by comparing the resultant MC 
correlation function with the given input $C_{2}$ function. The source 
size parameters were deduced from the experimental correlation function 
after applying the correction factors to the $C_{raw}$ function. 
The iteration ends when the extracted source-size parameters agree 
with the given parameters within an acceptable accuracy. 
The detailed procedure is described in the first article in 
Ref.~\cite{NA44} and more discussion can be found in Ref.~\cite{zajc84}.

\begin{figure}[b]
\label{hbt}
\epsfig{file=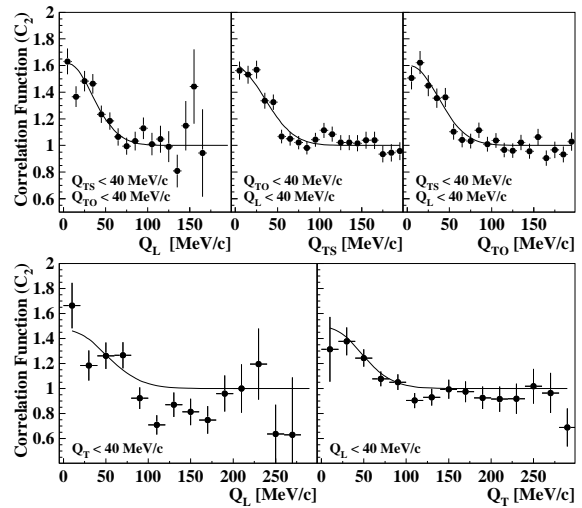,width=3.375in}
\caption{Projections of the $K^{+}K^{+}$ correlation functions at 
$p_{T}\approx 0.25$ GeV/{\it c} (top) and $p_{T}\approx 0.91$ GeV/{\it c} 
(bottom). The solid lines show the projections of the fit with the 
Gaussian parametrization.}
\label{fig:2}
\end{figure}

After the PID selection and quality cuts, we have around $20 \times 10^3$ pairs 
in both the horizontal and vertical modes at the lower $p_T$, 
while $17 \times 10^3$ pairs in the horizontal mode at the higher $p_T$. 
A three-dimensional fit is applied to the lower $p_T$ data. 
The observed momenta of each kaon pair are transformed to the 
LCMS, in which the momentum 
sum of two particles along the beam axis is zero, 
$\vec{p}_{z1}+\vec{p}_{z2}=0$.
The momentum variables of interest are defined in this system.  
The average momentum of the pair in the frame is
$\vec{k}_T$ = $(\vec{p}_{T1} + \vec{p}_{T2})$/2. 
$Q_L$ is a projection of the momentum difference onto the beam axis, and 
$Q_T$ is the momentum difference perpendicular to the beam axis. 
$Q_T$ is further divided into two components. $Q_{TO}$ is the
component of $Q_T$ along $\vec{k}_T$, and $Q_{TS}$ is the component 
of $Q_T$ perpendicular to both $\vec{k}_T$ and the beam axis.
The momentum resolution of $Q_{L}$, $Q_{TO}$, and $Q_{TS}$, 
including effects of multiple scattering in the target, were 
around 10, 20, and 20 MeV/{\it c}, respectively. 
The correlation functions $C_{2}$ 
in both the horizontal and vertical modes are simultaneously fitted 
using the maximum likelihood method with 
\begin{equation}
C_2(Q_{TO},Q_{TS},Q_L)=D(1+ \lambda e^{-Q_{TO}^2R_{TO}^2-
Q_{TS}^2R_{TS}^2-Q_{L}^2R_{L}^2}),  
\label{eq:C_2_3dim}
\end{equation}
where $\lambda$ is a factor introduced to express chaoticity of 
quantum states of the source and $R$'s are variables representing 
multidimensional radii of the system in question. $D$ is a free 
parameter for normalization in each mode. 
A two-dimensional equation replacing 
$-Q_{TO}^2R_{TO}^2-Q_{TS}^2R_{TS}^2$ in Eq.~\ref{eq:C_2_3dim} 
with $-Q_T^2R_T^2$ is employed to fit the correlation function in 
the horizontal mode at the higher $p_T$. 
The fit results are given in Table~I, where the systematic 
uncertainties reflect the effect of 
(i) cut parameters to define a track,
(ii) cut parameters to select pairs, 
(iii) momentum resolution, 
(iv) two-track resolution, 
(v) momentum distribution of particle production in MC, and
(vi) fitting to finite bins. 
Figure~\ref{fig:2} shows projections of the correlation function  
onto each axis of the momentum difference where the projection 
is over the lowest 40 MeV/{\it c} of the other directions in the 
momentum difference. In these plots, the solid lines show the 
results of fit projected in the same way as the data. 

\vspace{0.2in}
\noindent TABLE I.  Fit results of Gaussian parametrizations of the $K^{+}K^{+}$ 
correlation functions at low and high $p_{T}$. 
The errors are statistical and systematical ones.
\begin{tabular}{ccc}
\hline\hline
$<p_{T}>$ [GeV/{\it c}] & 0.25 & 0.91 \\
$\lambda$ & 0.84 $\pm$ 0.06 $\pm$ 0.07 & 0.61 $\pm$ 0.20 $\pm$ 0.16 \\
$R_L$ [fm] & 4.36 $\pm$ 0.33 $\pm$ 0.32 & 3.20 $\pm$ 0.54 $\pm$ 0.45 \\
$R_{T}$ [fm] & n/a & 3.59 $\pm$ 0.67 $\pm$ 0.97 \\
$R_{TS}$ [fm] & 4.04 $\pm$ 0.28 $\pm$ 0.32 & n/a \\
$R_{TO}$ [fm] & 4.12 $\pm$ 0.26 $\pm$ 0.31 & n/a \\
$\chi^2$/{\it d.o.f.} & 5139 /2978 & 117 / 107 \\
\hline\hline
\end{tabular}

\begin{figure}[tbph]
\label{mt}
\epsfig{file=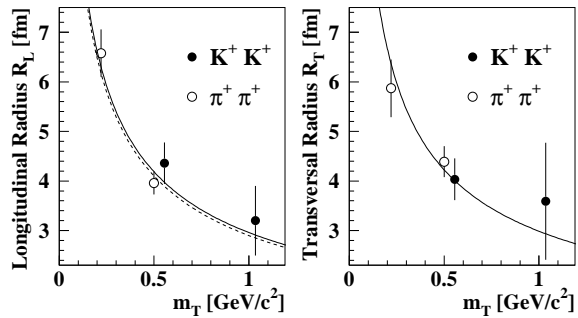,width=3.375in}
\caption{The longitudinal and transversal source-size parameters 
of $K^{+}K^{+}$ at $p_{T}\approx 0.25$ GeV/{\it c} 
and 0.91 GeV/{\it c} (solid circle), compared with those of pions 
(open circle). The dashed curve shows a fit to $R=A/\sqrt{m_{T}}$ 
for pions, while the solid curve is the fit to the pion and kaon 
data-points in each plot.}
\label{fig:3}
\end{figure}

The three source-size parameters ($R_L$, $R_{TO}$, $R_{TS}$) in the 
three-dimensional fit are quite similar to each other --- as was observed 
in pion correlation measurements in the same colliding 
system~\cite{PRC(John)}. 
Since $R_{TS}$ and $R_L$ represent the geometric information of the 
source most directly, they are compared in Fig.~\ref{fig:3} with 
those of pions. 
The present data at $m_T \approx 0.55$ GeV/${\it c}^2$ seems to stay 
on the $m_T$ scaling curve (dashed) which came from the pion correlation 
measurements. 
To confirm this tendency, we put $R_T$ and $R_L$ from the two-dimensional 
fit at the higher $p_T$ ($m_T \approx 1.0$ GeV/${\it c}^2$) on the same 
plot. They also seem consistent with the scaling curve. 
A fit to the four data points in each plot with a single scaling curve, 
$R=A/\sqrt{m_{T}}$, gives $A=3.0\pm 0.2~{\rm fm {GeV}^{1/2}}{\it
c^{\rm -1}}$ in both cases as shown with solid curves.

The dependences on $m_T$ are predicted with models including 
hydrodynamical expansion 
in a source~\cite{Akkelin95,Wiedemann96,Lorstad96}.
The experimental radius parameters are interpreted 
as a length of homogeneity which is in turn dependent on a geometrical 
source size $R_{geom}$ and a thermal length $R_{therm}$. 
The ``boost invariant expansion'' along the longitudinal 
direction leads to an expression $R_L=\tau\sqrt{T_0/m_{T}}$, 
ignoring the contribution from $R_{geom}$. 
Assuming a freeze-out temperature $T_{0}$ of 100--140 MeV, 
we could extract the freeze-out time $\tau$ from the present $A$ 
parameter to be 7--10 fm/{\it c}, which is in good agreement with 
the WA98~\cite{WA98} and NA49~\cite{NA49} results. 
In a hydrodynamic model under certain conditions~\cite{Lorstad96},
one can derive an analytic expression for the radii
$R_T\simeq R_{L}\simeq \tau\sqrt{T_0/m_{T}}$. Our data are consistent 
with such a scenario. The common $m_T$ scaling for pions and kaons 
may imply that thermal freeze-out occurs simultaneously for both pion 
and kaons and that therefore they receive a common Lorentz boost.
This is consistent with the hydrodynamic hypothesis. A similar 
conclusion can be drawn from the linear increase of the single 
particle inverse slopes with mass~\cite{PRL78}. 

We derived the duration time $\Delta \tau$ of kaon emission from 
quadratic difference of $R_{TO}$ and $R_{TS}$ in Eq.~(\ref{eq:C_2_3dim})
as $\Delta\tau=\sqrt{R_{TO}^2-R_{TS}^2}/\beta$, 
where $\beta$ is the transverse velocity of the kaon pair in the LCMS 
frame. We find 
$\Delta \tau = 2.2\pm5.2(stat)\pm 6.1(syst)$ fm/{\it c}. 
The kaon duration time is short and similar to those observed for 
pions in the same colliding system and for kaons in the S + Pb 
collisions.
The present result excludes simple scenarios of a prolonged mixed 
phase anticipated in a first-order phase transition from a QGP phase. 

The NA44 collaboration wishes to thank the staff of the CERN PS-SPS
accelerator complex for their excellent work. 
We are also grateful for support given by 
the Danish Natural Science Research Council;
the Japanese Society for the Promotion of Science; 
the Ministry of Education, Science and Culture, Japan; 
the Swedish Science Research Council;
the Austrian Fond f\"{u}r F\"{o}rderung der Wissenschaftlichen Forschung; 
the National Science Foundation, and
the US Department of Energy.

\end{document}